\documentclass[journal]{IEEEtran}

\usepackage[applemac]{inputenc}
\usepackage[american]{babel}
\usepackage{amsmath}
\usepackage{amssymb}
\usepackage{amsthm} 
\usepackage{thmtools}
\usepackage{cite} 
\usepackage{hyperref} 

\usepackage{pgfplots}

\usepackage{acronym}

\usepackage{balance}

\ifCLASSOPTIONcompsoc
\usepackage[caption=false,font=normalsize,labelfont=sf,textfont=sf]{subfig}
\else
\usepackage[caption=false,font=footnotesize]{subfig}
\fi

\usepackage{ctable}
\definecolor{green}{RGB}{34,150,34}
\newtheorem{definition}{Definition}

\newtheorem{proposition}{Proposition}
\newtheorem{example}{Example}

\begin{document}
	
\acrodef{FACSA}[FA-CSA]{frame asynchronous CSA}
\acrodef{FSCSA}[FS-CSA]{frame synchronous CSA}
\acrodef{FACSAUB}[FA-CSA-UB]{FA-CSA with uniform slot selection and boundary effect}
\acrodef{FACSAUNB}[FA-CSA-UNB]{FA-CSA with uniform slot selection and no boundary effect}
\acrodef{FACSAFB}[FA-CSA-FB]{FA-CSA with first slot fixed and boundary effect}
\acrodef{FACSAFNB}[FA-CSA-FNB]{FA-CSA with first slot fixed and no boundary effect}
\acrodef{FACSAU}[FA-CSA-U]{FA-CSA with uniform slot selection}
\acrodef{FACSAF}[FA-CSA-F]{FA-CSA with first slot fixed}
\acrodef{pmf}{probability mass function}
\acrodef{cdf}{cumulative density function}
\acrodef{SCCSA}[SC-CSA]{spatially-coupled CSA}

\newcommand{\uA}{\mathsf{A}}

\newcommand{\otoprule}{\midrule[\heavyrulewidth]}
\newcommand{\IID}{i.i.d.\ }	
\newcommand{\IE}{i.e., } 
\newcommand{\EG}{e.g., } 
\newcommand{\etal}{\emph{et al.\ }} 
\newcommand{\cp}{cp.\ } 
\newcommand{\figref}[1]{Fig.\,\ref{fig:#1}}
\newcommand{\figreff}[2]{Figs.\,\ref{fig:#1} and \,\ref{fig:#2}}
\newcommand{\RN}[1]{%
	\textup{\uppercase\expandafter{\romannumeral#1}}
}
\newcommand{\Rho}{\operatorname{P}} 
\newcommand\given[1][]{\:#1\vert\:} 
\newcommand {\define}{\stackrel{\Delta}{=}}
\newcommand{\Po}[1]{\ensuremath{\text{\textsf{Po}}(#1)}} 
\newcommand{\Binomial}[2]{\ensuremath{\text{\textsf{Binom}}\left(#1,#2\right)}}
\newcommand{\Prob}[1]{\ensuremath{\text{\textsf{Pr}}(#1)}}
\newcommand{\N}{n} 				
\newcommand{\NRX}{\N_{\text{RX}}} 
\newcommand{\T}{i} 				 
\newcommand{\G}{g}				
\newcommand{\SNR}{E_\mathrm{b}/N_0}
\newcommand{\Graph}{\mathcal{G}}
\newcommand{\EdgeG}{\mathcal{E}}
\newcommand{\Variables}{\mathcal{V}}
\newcommand{\Checks}{\mathcal{C}} 
\newcommand{\M}{M}				
\newcommand{\Muni}{M_{\mathrm{FA-U}}}				
\newcommand{\m}{m}				
\newcommand{\Mtime}{M_{\T}} 	

\newcommand{\tslot}{\tau}
\newcommand{\pii}{p_{i\rightarrow i}}
\newcommand{\PI}{p_i}
\newcommand{\PJ}{p_j}
\newcommand{\qi}{q_i}
\newcommand{\qiPJ}{q_{i+j}}
\newcommand{\PIPJ}{p_{i-j}}
\newcommand{\pij}{p_{i\rightarrow j}}
\newcommand{\qii}{q_{i\rightarrow i}}
\newcommand{\qij}{q_{i\rightarrow j}}
\newcommand{\qipj}{q_{j \rightarrow i}}
\newcommand{\pipj}{p_{k \rightarrow i}}
\newcommand{\qavi}{\tilde{q}_i}
\newcommand{\pavi}{\tilde{p}_i}
\newcommand{\CNddother}{\Rho^{i\rightarrow \Kset}}
\newcommand{\CNddotherr}{\Rho^{i\rightarrow \Kset}_{r_2}}
\newcommand{\CNedother}{\rho^{i\rightarrow \Kset}}
\newcommand{\CNedotherr}{\rho^{i\rightarrow \Kset}_{r_2}}
\newcommand{\CNddsame}{\Rho^{i\rightarrow i}}
\newcommand{\CNedsame}{\rho^{i\rightarrow i}}
\newcommand{\CNddsamer}{\Rho^{i\rightarrow i}_{r_1}}
\newcommand{\CNedsamer}{\rho^{i\rightarrow i}_{r_1}}
\newcommand{\CNedsamerprime}{\rho_{i\rightarrow i,r_1'-1}}
\newcommand{\CNdd}{\Rho }
\newcommand{\CNed}{\rho }
\newcommand{\VNdd}{\Lambda }
\newcommand{\VNed}{\lambda }
\newcommand{\VNedTilde}{\lambda^{i\rightarrow \Jset}}
\newcommand{\VSameTilde}{\lambda^{i\rightarrow i}}
\newcommand{\VNddOpt}{\Lambda^{\star}}

\newcommand{\PLR}{\bar{p}} 

\newcommand{\user}{u}	
\newcommand{\Sset}{\mathcal{S}} 
\newcommand{\numCNs}{\mu \left(\Sset\right)} 
\newcommand{\numVNs}{\nu \left(\Sset\right)} 
\newcommand{\numVNsL}{v_l\left(\Sset\right)} 
\newcommand{\numVNl}[1]{{v_{#1}\left(\Sset\right)}}
\newcommand{\numVNsD}{v_d\left(\Sset\right)} 
\newcommand{\iso}{c(\Sset)} 
\newcommand{\A}{\mathcal{A}} 
\newcommand{\Astar}{\mathcal{A}^{\star}} 
\newcommand{\ps}{\Pr\left(\user \in \Sset |m\right)} %
\newcommand{\profileS}{\ensuremath{\boldsymbol{v}(\Sset)}}
\newcommand{\aS}{a(\Sset,m)}
\newcommand{\aSFS}{a(\Sset,m)}
\newcommand{\bS}{b_{\mathrm{FA-F}}(\Sset)}
\newcommand{\dS}{d_{\mathrm{FA-F}}(\Sset)}
\newcommand{\bSFS}{b_{\mathrm{FS}}(\Sset)}
\newcommand{\dSFS}{d_{\mathrm{FS}}(\Sset)}
\newcommand{\aSFAuni}{a(\Sset,m)}
\newcommand{\bSFAuni}{b_{\mathrm{FA-U}}(\Sset)}
\newcommand{\dSFAuni}{d_{\mathrm{FA-U}}(\Sset)}
\newcommand{\fS}{f(\Sset,m)}
\newcommand{\delaymax}{\delta_{\text{max}}}
\newcommand{\psFAunitwo}[2]{\theta_{\text{FA-U}} \left(#1,#2\right)}
\newcommand{\psFAuni}[3]{\theta_{\text{FA-U}} \left(#1,#2,#3\right)}
\newcommand{\psFS}[2]{\theta_{\text{FS} }\left(#1,#2\right)}
\newcommand{\psM}{\theta \left(\Sset,\M\right)}
\newcommand{\Eps}{\mathbb{E}_{\M_\T} \left(\ps\right)}
\newcommand{\gstar}{\G^{\star}}
\newcommand{\gbound}{\G^{\star}_{\text{Bound}}}
\newcommand{\gstarfaBE}{\G^{\star}_{\text{FA-FB}}}
\newcommand{\gstarfaUB}{\G^{\star}_{\text{FA-UB}}}
\newcommand{\gstarfanoBE}{\G^{\star}_{\text{FA-FNB}}}
\newcommand{\gstarfanoUNB}{\G^{\star}_{\text{FA-UNB}}}
\newcommand{\gstarfs}{\G^{\star}_{\text{FS}}}
\newcommand{\lmax}{q(\Sset)}
\newcommand{\Jset}{\mathcal{J}_i}
\newcommand{\Kset}{\mathcal{K}_i}

\title{On Frame Asynchronous Coded Slotted ALOHA: Asymptotic, Finite Length, and Delay Analysis}
%
%
%

\author{Erik~Sandgren,
        Alexandre~Graell i Amat,~\IEEEmembership{Senior Member,~IEEE,}
        and~Fredrik~Br\"annstr\"om,~\IEEEmembership{Member,~IEEE}
        
\thanks{This paper was presented in part at the 9th International Symposium on Turbo Codes \& Iterative Information Processing, Brest, France, September 2016.}
\thanks{This work was partially funded by the Swedish Research Council under grants 2011-5950 and 2011-5961.}
\thanks{E. Sandgren was with the Department
of Signals and Systems, Chalmers University of Technology, SE-41296 Gothenburg, Sweden. He is now with Cybercom Group, Gothenburg, Sweden, (e-mail: erik.sandgren92@gmail.com).}
\thanks{A. Graell i Amat and F. Br\"annstr\"om are with the Department
of Signals and Systems, Chalmers University of Technology, SE-41296 Gothenburg, Sweden, (email: \{alexandre.graell, fredrik.brannstrom\}@chalmers.se).}
}

%
%

\markboth{}%
{}
%


\IEEEoverridecommandlockouts

\maketitle

\begin{abstract}
	We consider a frame asynchronous coded slotted ALOHA (FA-CSA) system for uncoordinated multiple access, where users join the system on a slot-by-slot basis according to a Poisson random process  and, in contrast to standard frame synchronous CSA (FS-CSA), users are not frame-synchronized. We analyze the performance of FA-CSA in terms of packet loss rate and delay.  In particular, we derive the (approximate) density evolution that characterizes the asymptotic performance of FA-CSA when the frame length goes to infinity. We show that, if the receiver can monitor the system before anyone starts transmitting, a boundary effect similar to that of spatially-coupled codes   occurs, which greatly improves the iterative decoding threshold. Furthermore, we derive tight approximations of the error floor (EF) for the finite frame length regime, based on the probability of occurrence of the most frequent stopping sets. We show that, in general, FA-CSA provides better performance in both the EF and waterfall regions as compared to FS-CSA. Moreover, FA-CSA exhibits better delay properties than FS-CSA.
\end{abstract}

\IEEEpeerreviewmaketitle
\section{Introduction}

\IEEEPARstart{U}{ncoordinated} multiple access is necessary in any communications system where coordinated resource allocation is not possible or too costly. 
Classical uncoordinated techniques include the ALOHA systems introduced in the 1970s \cite{abramson1970aloha,roberts1972aloha} and carrier sense based techniques \cite{kleinrock1975csma}, such as  carrier sense multiple access with collision avoidance.
To provide reliable communication, these techniques typically require a retransmission policy, which introduces the need for a separate feedback channel and incurs in a possibly large delay. 

Recently, a considerable interest for finding novel solutions to provide  reliable, low latency communication in dynamical systems has emerged, driven by the stringent requirements of the upcoming 5G communication systems \cite{METIS}. 
One promising uncoordinated multiple access technique is coded slotted ALOHA (CSA) \cite{casini2007contention,liva2011graph,paolini2015csa}. CSA builds on the classical slotted ALOHA technique and borrows ideas from error correcting codes to provide highly reliable uncoordinated multiple access. 

In CSA, transmission is organized into frames, each consisting of the same number of slots. Two key ingredients of CSA is to let users replicate each packet a number of times\footnote{{We would like to remark that the term CSA was coined in \cite{paolini2015csa} to describe a quite general scheme where users use an arbitrary error correcting code (not necessarily a repetition code as in the seminal papers\cite{casini2007contention,liva2011graph}) to encode their messages prior to transmission on the multiple access channel. In this paper, we consider repetition codes as in \cite{liva2011graph},  but with a slight abuse of language we use the more general term CSA.}} within a frame and to perform iterative decoding using successive interference cancellation (SIC) \cite{casini2007contention}. Furthermore, in standard CSA,  referred to as \ac{FSCSA} in the sequel, all users are assumed to be frame-synchronized.

A major contribution made in \cite{liva2011graph} was realizing that there exists a connection between CSA and codes on graphs. Indeed, a CSA system can be described by a bipartite graph and SIC can be performed over the graph similarly to decoding of graph-based codes. Therefore, it is not surprising that the performance of CSA resembles that of graph-based codes, \IE the packet loss rate (PLR) curve is characterized by a waterfall (WF) region for medium-to-high system loads and it shows an error floor (EF) for low system loads. Furthermore, in the asymptotic regime of infinite frame length, CSA exhibits a threshold behavior, which can be characterized using density evolution (DE). The DE for FS-CSA was derived in \cite{liva2011graph}.

To improve the delay performance of FS-CSA, a \ac{FACSA} system was proposed in \cite{meloni2012sliding}. In contrast to FS-CSA, users in FA-CSA are not frame-synchronized. Instead, a user that joins the system selects slots for its packet replicas from a number of subsequent slots, which form its local frame. The duration of the local frame (in number of slots) is the same for each user. Using simulations, it was shown in \cite{meloni2012sliding} that, in addition to improve the average delay, \ac{FACSA} outperforms \ac{FSCSA} in terms of throughput.

In this paper, we provide a thorough analysis of the PLR performance of \ac{FACSA} in the asymptotic and finite frame length regime.  More specifically, we derive the (approximate) DE that governs the asymptotic performance of \ac{FACSA} in the limit of very large frame lengths. Interestingly, if the receiver can monitor the system before transmitting, a boundary effect similar to that of spatially-coupled codes \cite{Jim99,Kud11} naturally arises. This greatly improves the iterative decoding threshold as compared to that of \ac{FSCSA}. Similar improved thresholds were achieved in \cite{liva2012spatially}, where the concept of spatial coupling was applied to CSA. However, while the spatially coupled structure is enforced by design in \cite{liva2012spatially}, in \ac{FACSA} it is inherent to the system. 

We also derive analytical approximations of the PLR in the finite frame length regime in order to predict the EF of \ac{FACSA}. The analysis is based on the framework introduced in \cite{ivanov2015error} and \cite{ivanov2015broadcast}, where the probability of occurrence of minimal stopping sets was used to approximate the PLR of \ac{FSCSA}. The EF of \ac{FSCSA} and \ac{FACSA} was also analyzed in \cite{del2013acrda,del2013framework}. However, the analysis is restricted to regular distributions and considers a single, trivial stopping set.
We show that in the finite frame length regime \ac{FACSA} yields
superior performance than that of \ac{FSCSA} in both the WF and EF regions. Furthermore, by means of simulations, we confirm and further elaborate on the results regarding the delay performance of FA-CSA in \cite{meloni2012sliding}. We show that \ac{FACSA} yields lower average delay than \ac{FSCSA}, albeit at the cost of a higher maximum delay. We then compare \ac{FACSA} and \ac{FSCSA} with a constraint on the maximum delay and receiver memory size, and show that \ac{FACSA} achieves better PLR than \ac{FSCSA} also in this case. Finally, we compare \ac{FACSA} and \ac{SCCSA} in the finite frame length regime. It is shown that \ac{FACSA} performs better than \ac{SCCSA}, both in terms of PLR and delay.

We would like to remark that the boundary effect that arises in FA-CSA has already been observed for other asynchronous random access techniques in the past. In particular, in \cite{Pfi14} it was recognized that the excellent performance of the interference cancellation system proposed in \cite{Hou06} for the CDMA2000 1xEV-DO reverse link was due to spatial coupling. At the time of \cite{Hou06}, however, the concept of spatial coupling was not known yet.

The remainder of this paper is organized as follows. The system model and the bipartite-graph representation of CSA are presented in Section~\ref{sec:system_model}. In Section~\ref{sec:DD}, the degree distributions of FA-CSA are derived. The DE equations of FA-CSA are then derived in Section~\ref{sec:DE} using the analysis in Section~\ref{sec:DD}.
The analysis of the performance of FA-CSA in the finite frame length regime is addressed in Section~\ref{sec:EF}. Finally, Section~\ref{sec:results} presents and discusses numerical results and Section~\ref{sec:conclusion} concludes the paper.

\section{System Model}
\label{sec:system_model}
\begin{figure*}[]
	\centering
	\includegraphics[width=2\columnwidth]{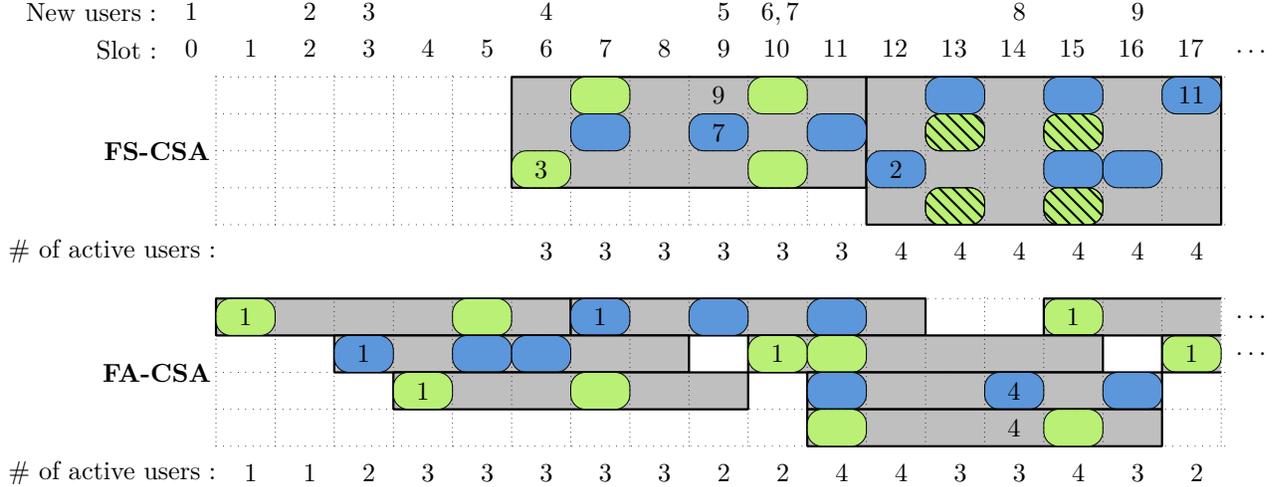}
	\vspace{-2ex}
	\caption{An illustration of \acs{FSCSA} and \acs{FACSAFB}. Both systems have the same new users joining. Green and blue slots represent replicas of degree-$2$ and degree-$3$ users, respectively. The four striped slots constitute a stopping set and cannot be resolved by the iterative decoder. }
	\label{fig:examp}
	\vspace{-2ex}
\end{figure*}

We consider a CSA system with multiple users transmitting to a common receiver, where time is divided into slots, each of duration $\tslot$, and where users are slot-synchronized.  A user that joins the system generates a message and selects a repetition factor $l$ randomly according to a predefined degree distribution \cite{liva2011graph}. The message is then mapped into a physical layer packet of duration $\tslot$ (including guard intervals), such that one packet can be sent within one slot. The user transmits $l$ copies (called replicas) of the packet in randomly selected slots.  A user that repeats its packet $l$ times is called a degree-$l$ user and similarly a slot containing $r$ replicas from different users is called a degree-$r$ slot. It is assumed that a packet can always be decoded if at least one of its replicas is in a degree-$1$ slot. Furthermore, to facilitate decoding, each replica of a packet contains pointers to all other replicas of that packet. Throughout the paper we assume that perfect SIC can be performed to remove the interference of a user's replicas once its packet has been decoded. This assumption makes the analysis of the system more feasible. In \cite{casini2007contention,liva2011graph} actual (low complexity) SIC was implemented with little performance degradation as compared to perfect SIC.

We assume that users join the system on a slot-by-slot basis according to a Poisson process and let $K$ denote the number of users that join in a slot. Therefore, $K$ is a Poisson-distributed random variable (RV) with mean $\G$, $K \sim \Po{\G}$, where  $\G$ is the average system load in users per slot. The probability that $k$ users join the system in a given slot is thus
\begin{equation}
	\Pr(K=k) =  \frac{e^{-\G}\G^k}{k!}.
\end{equation}
This is a common user model for multiple access techniques, used, \EG  in the original ALOHA systems \cite{abramson1970aloha,roberts1972aloha}.

We define the two most important performance measures for CSA.
\begin{definition}
	The PLR, denoted by $\PLR$, is the average probability that the packet of an arbitrary user is never resolved. 
	\label{def:plr}
\end{definition}
\begin{definition}
	\label{def:delay}
	The delay of a resolved user's packet is the number of slots between the slot the user joins the system and the slot following the decoding of its packet. 
\end{definition}

\subsection{Frame Synchronous Coded Slotted ALOHA}

In \ac{FSCSA}, communication takes place during \textit{global frames} consisting of $\N$ slots each. A degree-$l$ user that joins the system waits until the next frame and transmits its $l$ replicas in randomly chosen slots of that frame. We then say that the user is \emph{active} during the whole duration of the frame. We denote by $\M \sim \Po{\N\G}$ the RV representing the number of active users per frame. Note that the active users in a frame are all the users that joined the system during the previous frame.

We consider decoding of \ac{FSCSA} performed on a slot-by-slot basis. Assume the decoding of slot $i$. First, the interference caused by packets for which replicas in previous slots have already been decoded is canceled from the slot. The receiver then checks if slot $i$ is a degree-$1$ slot and, if not, the decoding of slot $i$ is stopped. Otherwise, the packet in slot $i$ is decoded and the interference from all its replicas canceled from the corresponding past slots.  The receiver then proceeds to iteratively find any degree-$1$ slots in its memory, decode the packets in these slots, and cancel the interference of all replicas of the decoded packets. This process continues until no new degree-$1$ slots are found or a maximum number of iterations is reached. 

\subsection{Frame Asynchronous Coded Slotted ALOHA}

In \ac{FACSA}, when a degree-$l$ user joins the system it waits until the next slot and transmits a first replica in that slot. This enforces users to be slot-synchronous. The remaining $l-1$ replicas are distributed uniformly within the $\N-1$ subsequent slots. Similarly to \ac{FSCSA}, $\N$ is the frame length of \ac{FACSA}. However, contrary to \ac{FSCSA}, slots are not arranged in global frames. We say that a user is active during the $\N$ slots following the slot where it joins the system and, accordingly, we call the $\N$ slots where a user is active its \textit{local frame}. Note that a user is not active in the slot where it joins the system, but only during the $n$ slots it can transmit in. Because a user always transmits in the first slot of its local frame, we call this system \ac{FACSAF}. 

Decoding of \ac{FACSA} is performed in a similar way as for \ac{FSCSA}, with the only difference that the receiver needs to consider not only the slots of a current frame, but all slots of the entire history of the system. In practice, the receiver cannot consider infinitely many slots and has a finite memory.  We denote by $\NRX$ the size of the receiver memory in number of slots. A finite $\NRX$ creates the notion of a \textit{sliding-window} decoder. It was shown in \cite{meloni2012sliding} that increasing $\NRX$ beyond $5\N$  does not improve performance in general. 

We let
\begin{equation}
\Mtime \sim \Po{\mu_i}
\end{equation}  
denote the number of active users in the $i$th slot, which is Poisson-distributed with mean $\mu_i$.  The active users in slot $i$ are all users that joined the system in slots $[i-n,i-1]$.  We  consider two different models for the initialization of the system, \IE for  $1\leq \T \leq \N$. The first model assumes that there are no active users at $\T=0$. In this case
\begin{equation}
\mu_i = \begin{cases} 
\T\G & \text{ for } 1 \leq \T <\N \\
\N\G & \text{ for } \T \geq \N
\end{cases},
\label{eq:init1}
\end{equation}
and we say that a \textit{boundary effect} is present for this model. Effectively, this means that the $\N -1$ first slots of the system have lower average degree than the rest. 

The second model assumes that there are already $\M \sim \Po{\N\G}$ active users at $\T=0$.\footnote{We remark that for the computation of the PLR, all users transmitting in or after $\T = 0$ are considered. Note that the users already active at slot $i=0$ will have a worse PLR than users joining at or after $i=0$, since all of their replicas may not be available. Furthermore, users already active at $i = 0$ will affect the PLR of users joining at $i \geq 0$ by causing further interference.} Thus,
\begin{equation}
\mu_i = \N\G \quad \text{ for all } \T \geq 1.
\label{eq:init2}
\end{equation}
For this model, all considered slots have the same average degree. 

The system with boundary effect corresponds to a system where the receiver is present at the very start of the communication or, potentially,  a system with periods of low load.  A good example is satellite networks, where the receiver, i.e., the satellite, is naturally present at the beginning of the communication. Another example is road side infrastructures in a vehicular network as the intended receivers. On the other hand, the model with no boundary effect is useful for systems where the receiver joins an already ongoing communication, \EG a vehicle in an all-to-all broadcast vehicular network where vehicles exchange messages between each other, as considered in \cite{ivanov2015broadcast}. A vehicle will join and leave local networks with ongoing communication as it is moving. For the same reason, this model is also practical for devices in a device-to-device communication network.

In addition to the initialization models described above, we introduce a second model for the selection of slots for transmission aside from that of \ac{FACSAF}, where a degree-$l$ user selects all $l$ slots for transmission randomly from the  local frame. We call this system \ac{FACSAU}. \ac{FACSAU} is more similar to \ac{FSCSA} and provides a simplified analysis in some cases. However, we remark that \ac{FACSAF} is more practical and, as it is shown later, performs better in general. Therefore, the main focus of this paper is on \ac{FACSAF}.

In all, we consider four models for \ac{FACSA}, \IE \ac{FACSAFB}, \ac{FACSAFNB}, \ac{FACSAUB}, and \ac{FACSAUNB}. The terminology ``boundary effect'' will become clearer in Sections~\ref{sec:DD} and \ref{sec:DE}.
\subsection{Bipartite Graph Representation}
An instance of a CSA system can be completely represented by a bipartite graph  $\Graph=\{\Variables,\Checks,\EdgeG\}$, where $\Variables$ is the set of variable nodes (VNs), $\Checks$ is the set of check nodes (CNs), and $\EdgeG$ is the set of edges connecting the VNs and CNs. VNs represent users and CNs represent slots. There is an edge $e_{i \rightarrow j} \in \EdgeG$ from VN $i$ to CN $j$ if user $i$ transmits a replica in slot $j$. Decoding of CSA can be viewed as message passing on the underlying graph \cite{liva2011graph}.  The degree of a node is equal to the number of edges incident to the node.

\begin{example}
	An example of \ac{FSCSA} and \ac{FACSAFB} is depicted in \figref{examp}, with $\N=6$, $\G=0.5$, and where users select degree 2 or 3 with equal probability. Gray areas show the frames of \ac{FSCSA} and local frames for each user in \ac{FACSAFB}. Slots filled with green represent packets of degree-2 users and slots filled with blue represents packets of degree-3 users. The delay of each user is indicated by a number in the slot in which it is decoded. Furthermore, the four striped green slots in the second frame of the \ac{FSCSA} example collide in such a way that the packets in these slots can not be decoded. Such collision patterns are called \textit{stopping sets}.
\end{example}

\begin{figure}[]
	\centering
	\includegraphics[width=\columnwidth]{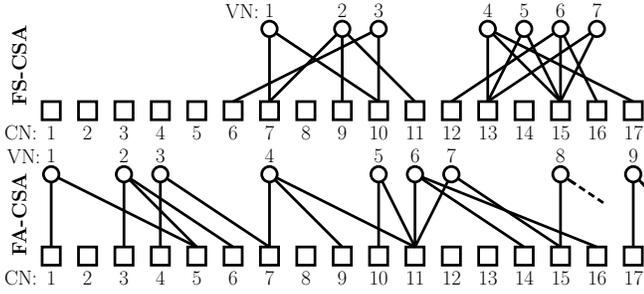}
	\vspace{-4ex}
	\caption{Equivalent graph representation of the systems depicted in \figref{examp}. VNs are represented by circles and CNs by squares.}
	\label{fig:bip_examp}
	\vspace{-3ex}
\end{figure}

 In \figref{bip_examp}, the graph representation for the two scenarios (FS-CSA and FA-CSA) in \figref{examp} is depicted.
 
\section{Degree Distributions}
\label{sec:DD}
In this section, we derive the VN degree and CN degree distributions for \ac{FACSA}.  We define the \textit{node-perspective} VN degree and CN degree distributions as 
\begin{equation}
	\label{eq:NPdegreedist}
	\VNdd(x) \define \sum_{l}^{} \VNdd_l x^l\quad \text{and } \quad  	\CNdd(x) \define \sum_{r}^{} \CNdd_r x^r,
\end{equation}
respectively, where $\VNdd_l$ is the probability that an arbitrary VN has degree $l$ and $\CNdd_r$ is the probability that a CN has degree $r$. $\VNdd(x)$ is under the control of the system designer and is subject to optimization. We introduce also the \textit{edge-perspective} VN degree and CN degree distributions  
\begin{equation}
\VNed(x)  \define \sum_{l}^{}\VNed_l x^{l-1}\quad \text{and } \quad   \CNed(x) \define \sum_{r}^{} \CNed_r x^{r-1}, 
\end{equation}
where $\VNed_l$ denotes the probability that an  edge is connected to a degree-$l$ VN and $\CNed_r$ denotes the probability that an  edge is connected to a degree-$r$ CN. The probabilities $\VNed_l$ and $\CNed_r$ are given by
\begin{equation}
	\VNed_l = \frac{l \VNdd_l}{\sum_{d}^{}d \VNdd_d } \quad \text{and } \quad  \CNed_r = \frac{r \CNdd_r}{\sum_{d}^{} d \CNdd_d},
\end{equation}
\IE $\VNed(x)=\VNdd'(x)/\VNdd' (1)$ and $\CNed(x) = \CNdd' (x)/ \CNdd' (1)$, where $f'$ denotes the derivative of the function $f$.

\subsection{Frame Asynchronous CSA with First Slot Fixed}
\label{sec:FACSA_first}

\begin{figure}
	\centering
	\includegraphics[scale=0.4]{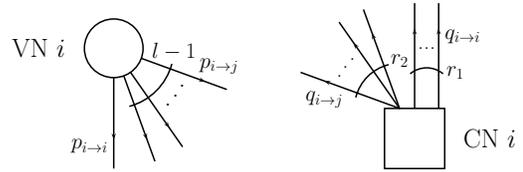}
	\vspace{-2ex}
	\caption{Class-$i$ VN and CN and their corresponding connectivity.} 
	\label{fig:DE_description}
	\vspace{-3ex}
\end{figure}

For FA-CSA with boundary effect, the first $\N$ CNs all have distinct degree distributions. This gives rise to different \textit{classes} of CNs and VNs. We call a CN at position $i$ (slot $i$) a class-$i$ CN. Similarly, a VN at position $i$ is a class-$i$ VN. Additionally, in \ac{FACSAF} a degree-$l$ class-$i$ VN always has one connection to a class-$i$ CN, \IE a fixed edge, and it has $l-1$ connections to randomly selected CNs of classes
\begin{equation}
	\Jset \define [i+1,i+n-1]. \label{eq:J}
\end{equation}
 The node connectivity for class-$i$ VNs and CNs of \ac{FACSAF} is depicted in \figref{DE_description}. Accordingly, we define the node-perspective VN degree distributions for \ac{FACSAF}
\begin{align}
\Lambda^{i\rightarrow i}(x)&= x, \nonumber\\ 
\Lambda^{i\rightarrow \Jset}(x)&=\sum_l\Lambda^{i\rightarrow \Jset}_lx^{l}\stackrel{(a)}{=}\sum_l\Lambda_lx^{l-1},
\label{eq:DistiJi}
\end{align} 
where $\Lambda^{i\rightarrow i}(x)$ represents the fixed connection, $\Lambda^{i\rightarrow \Jset}_l=\Lambda_{l+1}$ is the probability that a class-$i$ VN has $l$ connections to CNs of classes in $\Jset$, and in $(a)$ we made the change of variables $l \rightarrow l-1$. The corresponding edge-perspective degree distributions are 
\begin{align}
\VSameTilde(x) &= 1,\nonumber \\
\VNedTilde(x) &= \frac{\left(\Lambda^{i\rightarrow \Jset}\right)' (x)}{\left(\Lambda^{i\rightarrow\Jset}\right)' (1)}=\sum_l \lambda^{i\rightarrow \Jset}_l x^{l-2}, \label{eq:VNedTilde}
\end{align}
with $\lambda^{i\rightarrow \Jset}_l= \VNdd_l(l-1)/\sum_{l}^{}\VNdd_l(l-1)$. 
On the other hand, a class-$i$ CN is connected to $r_1$ class-$i$ VNs and to $r_2$ VNs of classes in the range 

\begin{equation}
\Kset \define \begin{cases} 
    \emptyset   & \text{for} \quad  i =1\\ 
	[1,i-1]     & \text{for} \quad  2 \leq i < n\\
	[i-n+1,i-1] & \text{for} \quad  i \geq n
\end{cases}.  \label{eq:K}
\end{equation}

Correspondingly, we define the CN degree distributions for \ac{FACSAF}
\begin{equation}
\CNddsame (x)=\sum_{r_1}\CNddsamer x^{r_1}\quad \text{and} \quad \CNddother (x)=\sum_{r_2} \CNddotherr x^{r_2},
\end{equation} where $\CNddsamer$ is the probability that a class-$i$ CN has $r_1$ edges incident to class-$i$ VNs, and $\CNddotherr$ is the probability that a class-$i$ CN has $r_2$ connections to VNs of classes in $\Kset$. The corresponding edge-perspective degree distributions are 
\begin{equation}
\rho^{i\rightarrow i}(x)= \frac{\left(\Rho^{i\rightarrow i}\right)'(x)}{\left(\Rho^{i\rightarrow i}\right)'(1)}= \sum_{r_1}\CNedsamer x^{r_1-1} 
\end{equation}
and
\begin{equation}
\rho^{i\rightarrow \Kset}(x)=\frac{\left(\Rho^{i\rightarrow \Kset}\right)'(x)}{\left(\Rho^{i\rightarrow \Kset }\right)'(1)}=\sum_{r_2}\CNedotherr x^{r_2-1}.
\end{equation}

\begin{proposition}
	\label{lemma:1}
	The class-$i$ CN degree distributions for \ac{FACSAF} are given by  
	\begin{equation}
	\CNddsame(x)=\rho^{i\rightarrow i}(x)= \exp(-\G(1-x))
	\end{equation}
	and
	\begin{equation}
	\CNddother(x) = \rho^{i\rightarrow \Kset}(x)=\exp\left(-\frac{\delta_i(\VNdd'(1)-1)}{n-1} (1-x)\right),
	\end{equation}
	with  
	\begin{equation}
	\delta_i = \begin{cases}
	\min(i-1,n-1)\G  & \text{for} \quad  \text{\ac{FACSAFB}} \\
	(n-1)\G  &							\text{for}  \quad \text{\ac{FACSAFNB}}
	\end{cases}	.
	\label{eq:deltai}
	\end{equation}
\end{proposition}
\begin{IEEEproof}
	Denote by $R_1$  the RV representing the number of edges connecting a class-$i$ CN to class-$i$ VNs. Clearly, $R_1 \sim \Po{\G}$, because each class-$i$ VN is connected through a single edge to the class-$i$ CN. Therefore, 
	\begin{equation}
	\CNddsamer =  \Pr(R_1=r_1) =\exp(-\G)\frac{\G^{r_1}}{r_1!}.
	\end{equation}
	Now $\CNddsame (x)$ is given by
	\begin{align}
		\label{eq:lemma1eq5}
		\CNddsame (x)  &= \sum_{r_1}^{}\CNddsamer x^{r_1} \nonumber\\
					  &=\sum_{r_1=0}^{\infty} \exp(-\G) \frac{\G^{r_1}}{r_1!} x^{r_1} \nonumber \\ 
					  &\stackrel{(a)}{=}\exp\left(-\G(1-x)\right),
	\end{align}	
	where in $(a)$ we used that $\sum_{n=0}^{\infty}\frac{x^n}{n!}=\exp(x)$. Furthermore, 
	\begin{align}
		\label{eq:lemma1eq6}
		\rho^{i\rightarrow i}(x)  &=\frac{\left(\Rho^{i\rightarrow i}\right)'(x)}{\left(\Rho^{i\rightarrow i }\right)'(1)}= \frac{\G \exp(-\G(1-x))}{\G \exp(0)} \nonumber\\
		                          &= \exp(-\G(1-x)).
	\end{align}	
We now denote by $R_{2,i}$ the number of edges connecting a class-$i$ CN to VNs of classes in the range $\Kset$, as given in \eqref{eq:K}. The number of VNs  
in $\Kset$ is  a Poisson RV, denoted by $K_i$, with mean $\delta_i$ given in \eqref{eq:deltai}. Each VN in $\Kset$ connects to the class-$i$ CN  with probability 
\begin{equation}
	z=\frac{\VNdd'(1)-1}{\N-1}.
	\label{eq:p_lemma_1}
\end{equation}
Applying the law of total probability this gives
\begin{align}
		\label{eq:lemma1part1}
	\CNddotherr &= \sum_{k=r_2}^{\infty}\Pr(R_{2,i}=r_2 \given K_i=k) \Pr(K_i=k) \nonumber\\ 
				&= \sum_{k=r_2}^{\infty}\binom{k}{r_2} z^{r_2} \left(1-z\right)^{k-r_2} \exp(-\delta_i) \frac{\delta_i^k}{k!} \nonumber \\
				&= \exp(-\delta_i)\left(\frac{z}{1-z}\right)^{r_2} \sum_{k=r_2}^{\infty} \frac{k!}{r_2! (k-r_2)!} \left(1-z\right)^{k}\frac{\delta_i^k}{k!}\nonumber \\
				&= \frac{\exp(-\delta_i)}{r_2!}\left(\frac{z}{1-z}\right)^{r_2} \sum_{k=r_2}^{\infty}\frac{((1-z)\delta_i)^k}{(k-r_2)!} \nonumber\\
				&\stackrel{(a)}{=} \frac{\exp(-\delta_i)}{r_2!}\left(z\delta_i\right)^{r_2} \sum_{k=0}^{\infty}\frac{((1-z)\delta_i)^k}{k!}\nonumber\\ 
				&\stackrel{(b)}{=} \exp(-z\delta_i) \frac{(z\delta_i)^{r_2}}{r_2!}.	 	
\end{align}
where in $(a)$ we used $k'=k-r_2$ and $k' \leftarrow k$ and in $(b)$ we used that $\sum_{n=0}^{\infty}\frac{x^n}{n!}=\exp(x)$.

Following similar steps as in \eqref{eq:lemma1eq5} and \eqref{eq:lemma1eq6}  gives
\begin{align}
	\CNddother(x) &= \rho^{i\rightarrow \Kset}(x) = \exp(-z\delta_i(1-x))\nonumber\\
														&=  \exp\left(-\frac{\delta_i(\VNdd'(1)-1)}{n-1} (1-x)\right),
\end{align}
where $z$ is given in \eqref{eq:p_lemma_1}.
\end{IEEEproof}

\subsection{Frame Asynchronous CSA with Uniform Slot Selection}
For \ac{FACSAU}, we need to consider only one degree distribution per CN class, as defined in \eqref{eq:NPdegreedist}. We denote by $\CNdd_i (x)$ and $\CNed_i (x)$ the node-perspective and edge-perspective class-$i$ CN degree distributions, respectively. 
\begin{proposition}
	The class-$i$ CN degree distribution for \ac{FACSAU} is given by  
	\begin{equation}
	\CNdd_i (x) =\CNed_i(x)= \exp\left(-\frac{\mu_i}{\N} \VNdd' (1)(1-x)\right),
	\end{equation}
	where for \ac{FACSAUB}  $\mu_i$ is given by \eqref{eq:init1} and for \ac{FACSAUNB} $\mu_i$ is given by \eqref{eq:init2}.
\end{proposition}
\begin{IEEEproof}
	A  class-$i$ CN can be connected with any of the $\Mtime$ VNs at position $i$. The probability that each of the $\Mtime$ VNs connects to the class-$i$ CN is $\VNdd' (1) / \N$. This setup is similar to the setup for the derivation of $\CNddother(x)$ in Proposition~\ref{lemma:1}. Taking similar steps, it directly follows that
		\begin{equation}
		\CNdd_i (x) =\CNed_i(x) = \exp\left(-\frac{\mu_i}{\N} \VNdd' (1)(1-x)\right).
		\end{equation}
\end{IEEEproof}
 
 The CN degree distribution for \ac{FSCSA}, derived in \cite{liva2011graph}, holds also when a Poisson user model is used, although the assumption that $\N \rightarrow \infty$ is not necessary. In fact, the CN degree distributions for \ac{FACSAUNB} and \ac{FSCSA} are the same, which is  expected since \ac{FACSAU}  is similar to  \ac{FSCSA} in that the edges of an arbitrary VN are connected to the CNs of its local frame the same way as a VN of \ac{FSCSA} connects edges to the CNs of a global frame. 
\section{Density Evolution Analysis}
\label{sec:DE}
In the asymptotic regime, \IE when $\N \rightarrow \infty$, CSA exhibits a threshold behavior: all users can be resolved if the system operates below a given system load, called threshold and denoted by $\gstar$. The threshold can be obtained via DE. In this section, we derive the DE equations for \ac{FACSA} with boundary effect.
\subsection{Density Evolution for Frame Asynchronous CSA with First Slot Fixed}
Because a class-$i$ VN is always connected to a class-$i$ CN, all edges of \ac{FACSAF} are not equivalent, see \figref{DE_description}. Therefore, we must differentiate between \textit{edge types}, and thus update $\pii$, $\pij$, $\qii$, and $\qij$ separately in the DE, where $\pii$ denotes the erasure probability from a class-$i$ VN to a class-$i$ CN, $\pij$ denotes the erasure probability from a class-$i$ VN  to a class-$j$ CN with $j \neq i$, $\qii$ denotes the erasure probability from a class-$i$ CN to a class-$i$ VN and $\qij$  denotes the erasure probability from a class-$i$ CN o to a class-$j$ VN with $j \neq i$. In the following, we derive the erasure probabilities $\pii$, $\pij$, $\qii$, and $\qij$.

We first derive $\pii$ and $\pij$. An outgoing message from a VN is in erasure if all incoming messages are in erasure. Consider first a class-$i$ VN of degree $l$, and call it VN $\uA$. VN $\uA$ has one edge connected to a class-$i$ CN and $l-1$ edges connected to CNs of classes $\Jset\triangleq[i+1,i+n-1]$ (see also Section~\ref{sec:FACSA_first} and Fig.~\ref{fig:DE_description}). Let $e_0$ be the edge connecting VN $\uA$ to a class-$i$ CN and let $e_1,e_2,\ldots,e_{l-1}$ be the edges connecting VN $\uA$ to CNs in $\Jset$.
Then, the probability that an outgoing message along the edge $e_0$ is in erasure is the probability that all incoming messages to VN $\uA$ along the edges $e_1,e_2,\ldots,e_{l-1}$ are in erasure. Now, since the edges connecting VN $\uA$ to the CNs of classes $\Jset$ ($|\Jset|=n-1$) are drawn uniformly at random, the average incoming erasure probability along edge $e_k$ is equal for all $k=1,\ldots,l-1$ and is given by
\begin{equation}
	\qavi = \frac{1}{\N - 1}\sum_{j \in \Jset}^{} \qipj,
	\label{eq:qavi}
\end{equation}
i.e., the average erasure probability of incoming messages from CNs in $\Jset$.
Thus, the probability that an outgoing message along the edge $e_0$ is in erasure is $p_{e_0}=\qavi^{l-1}$. Finally, averaging over the node-perspective VN degree distribution $\Lambda^{i\rightarrow \Jset}(x)$, $\pii$ is obtained as
\begin{align}
\label{eq:pii}
\pii &= \sum_{l}^{}\VNdd_l \qavi^{\,l-1} \stackrel{(a)}{=} \Lambda^{i\rightarrow \Jset}(\qavi),
\end{align}
where $(a)$ follows from \eqref{eq:DistiJi}.

In a similar way, $\pij$ can be derived using \eqref{eq:VNedTilde} as
\begin{align}
\pij &=  \qii\sum_{l}^{} \VNedTilde_l \qavi^{\, l-2} =  \qii \VNedTilde \left( \qavi \right),
\label{eq:pij}
\end{align}
for $j\in\Jset$.

We now derive $\qii$ and $\qij$. An outgoing message sent along an edge emanating from a class-$i$ CN is in erasure if at least one of its incoming messages (on the adjacent edges of the CN except the one on which the outgoing message is sent) is in erasure. Equivalently, a message from a CN is not in erasure if none of the incoming $r_1+r_2-1$ messages are in erasure. Consider a class-$i$ CN of degree $l$, and call it CN $\uA$. CN $\uA$ has $r_1$ edges connected to class-$i$ VNs and $r_2$ edges connected to VNs of classes in the range $\Kset$, defined in \eqref{eq:K}. Then, the probability that an outgoing message along an edge $e$ connecting CN $\uA$ to a class-$i$ VN is not in erasure, $\breve{p}_e$, is the probability that none of the remaining $r_1-1$ edges connecting CN $\uA$ to a class-$i$ VN and none of the $r_2$ edges connecting CN $\uA$ to a VN in the range $\Kset$ are in erasure, i.e., $\breve{p}_e=(1-\pii)^{r_1-1}(1-\tilde{p}_i)^{r_2}$, where
\begin{equation}
	\pavi = \begin{cases}
	0  & \text{for} \quad  i=1 \\
	\sum_{k\in \Kset}^{} \pipj/(i-1) &			\text{for}  \quad 1<i<n\\
	\sum_{k\in \Kset}^{} \pipj/(n-1)	&			\text{for} \quad i\geq n 
	\end{cases}	
\end{equation}
is the average erasure probability of incoming messages from VNs in $\Kset$. The probability that an outgoing message along an edge $e$ connecting CN $\uA$ to a class-$i$ VN is in erasure is then $p_e=1-\breve{p}_e=1-(1-\pii)^{r_1-1}(1-\tilde{p}_i)^{r_2}$. Now, averaging $(1-\pii)^{r_1-1}$ over the edge-perspective CN degree distribution $\rho^{i\rightarrow i}(x)$ and $(1-\tilde{p}_i)^{r_2}$ over the node-perspective CN degree distribution $\CNddother(x)$, $\qii$ is obtained as\footnote{The fact of averaging $(1-\pii)^{r_1-1}$ over the corresponding edge-perspective CN degree distribution and $(1-\tilde{p}_i)^{r_2}$ over the corresponding node-perspective CN degree distribution is due to the fact that we are considering the outgoing message along one of the $r_1$ edges connecting the class-$i$ CN to a class-$i$ VN. Furthermore, note that the summation on $r_1$ starts from $1$ because the edge under consideration cannot be connected to a class-$i$ CN with degree $r_1=0$, since it has at least one incident edge from class-$i$ VNs (the edge under consideration itself).}
\begin{IEEEeqnarray*}{rCl}
	\qii &= 1 -  &\left(\sum_{r_1 = 1}^{\infty}\CNedsamer(1-\pii)^{r_1-1} \right) \cdot  \\
	&  &\left( \sum_{r_2=0}^{\infty} \CNddotherr (1-\pavi{})^{r_2} \right) \\
	&\stackrel{(a)}{=} 1- &\left(\sum_{r_1 = 0}^{\infty}\frac{e^{-\G}\G^{r_1}}{r_1!}(1-\pii)^{r_1} \right)  \CNddother(1-\pavi{}) \\
	&= 1 - &\CNddsame(1-\pii) \CNddother(1-\pavi{})\\
	&= 1-  &\exp\left(-\G\pii\right) \exp\left(-\frac{\delta_i(\VNdd'(1)-1)}{n-1} \pavi{}\right) \IEEEyesnumber
	\label{eq:qii}
\end{IEEEeqnarray*}
where in $(a)$ we used $\CNedsamer = \CNddsamer$, $r'_1=r_1-1$ and $r'_1\leftarrow r_1$.

Similarly, $\qij$ can be derived as 
\begin{IEEEeqnarray*}{rCl}
	\qij&= 1 - &\left(\sum_{r_1 = 0}^{\infty}\CNddsamer(1-\pii)^{r_1} \right)\cdot \\
	&  &\left( \sum_{r_2=1}^{\infty} \CNedotherr (1-\pavi{})^{r_2-1} \right) \\
	&= 1-  &\exp\left(-\G\pii\right) \exp\left(-\frac{\delta_i(\VNdd'(1)-1)}{n-1} \pavi{}\right) \IEEEyesnumber
	\label{eq:qij}
\end{IEEEeqnarray*}
for $j \in \Kset$.
Note that $\qii=\qij$, which follows from the fact that $\CNddsame(x)=\CNedsame(x)$ and $\CNddother(x)=\CNedother(x)$ which, in turn, follows from the properties of the Poisson distribution  (see Proposition~\ref{lemma:1}). For a general user model, however, $\qii\neq\qij$.

DE is now performed by iteratively updating \eqref{eq:pii}--\eqref{eq:qij}, with $\pii,\,\pij,\,\qii$, and  $\qij$  initialized to $1$. The PLR of position $i$ can be computed as $\PLR_i=\VNdd(\qavi)\qii/\qavi$ and the threshold $\gstar$ is found by searching for the largest value of $g$ for which $\PLR_i$ converges to 0 for all positions. For a system without boundary effect, \eqref{eq:pii}--\eqref{eq:qij} are updated only for indices $i>n$.

We remark that exact DE requires $\N \rightarrow \infty$. This would require to keep track of an infinite number of node classes, which is unfeasible in practice. Therefore, the thresholds computed in Section~\ref{sec:results} must be seen as \emph{approximate} DE thresholds. However, we have found that it is sufficient to set $\N \approx 100$ and run DE over a chain of $20\N$ positions in order to obtain $\gstar$ with good precision. Considering larger values of $n$ does not change the obtained thresholds.
\subsection{Density Evolution for Frame Asynchronous CSA with Uniform Slot Selection}
For \ac{FACSAU} all edges are equivalent and therefore we do not need to consider different edge types. We denote by $\PI$ the probability that an erasure message is passed from a class-$i$ VN and by $\qi$ the probability that an erasure message is passed from the class-$i$ CN. 
It follows that
\begin{equation}
\PI =  \sum_{l}^{}\VNed_l\qavi^{\, l-1} = \VNed(\qavi),
\label{eq:DE1}
\end{equation}
where 
\begin{equation}
\qavi = \frac{1}{n}\sum_{j=i}^{i+n-1} \qi
\end{equation}
is the average erasure probability of the incoming messages to a class-$i$ VN.
Furthermore,
\begin{align}
	\qi &= 1- \sum_{r}^{} \CNed_{i,r} (1-\pavi)^{\, r-1} = 1- \CNed_i(1-\pavi) \nonumber \\
	    &= 1- \exp\left(-\frac{\mu_i}{\N} \VNdd' (1) \pavi \right),
\end{align}
where 
\begin{equation}
\pavi = \begin{cases}
\sum_{j=1}^{i} \PJ/i &			\text{for}  \quad 1\leq i < n\\
\sum_{j=i-n+1}^{i} \PJ/n	&			\text{for} \quad i\geq n 
\end{cases}	
\label{eq:DE4}
\end{equation}
is the average erasure probability of the incoming messages to the class-$i$ CN.

DE is performed similarly to \ac{FACSAF} by iteratively updating \eqref{eq:DE1}--\eqref{eq:DE4}, with $\PI$ and $\qi$ initialized to $1$. The PLR of position $i$ can be computed as $\PLR_i=\VNdd(\qavi)$.

\section{Finite Frame Length Analysis}
\label{sec:EF}
In the finite frame length regime, CSA exhibits an error floor in its PLR performance for low-to-medium loads $\G$. The EF is due to stopping sets, \IE graph substructures which make the iterative decoder fail. 
\begin{definition}
	A stopping set $\Sset$ is a connected bipartite subgraph with all CNs of degree strictly larger than 1.
\end{definition}
The EF of CSA is dominated by \textit{minimal stopping sets}.
\begin{definition}
	A minimal stopping set is a stopping set that does not contain a nonempty stopping set of smaller size.
\end{definition}
In this section, we find estimates of the EF by approximating the probability of occurrence of  minimal stopping sets. We first introduce some useful notation for a stopping set $\Sset$. Let $\numCNs$ denote the number of CNs, $\numVNs$  the number of VNs, and $\numVNsL$  the number of degree-$l$ VNs in $\Sset$. Moreover, we define the degree profile of a stopping set as the vector $\profileS=[\numVNl{0},\numVNl{1}, \ldots ,\numVNl{\numCNs}]$, and denote by $\iso$ the number of graph isomorphisms of $\Sset$ \cite[p.4]{bondy1976graph}. Unfortunately, there is no straightforward analytical expression for $\iso$. However, $\iso$ is tabulated in \cite[Table \RN{1}]{ivanov2015broadcast} along with $\numVNs$, $\numCNs$, and $\numVNsL$ for all $31$ minimal stopping sets of \ac{FSCSA} with $\numCNs\le 4$. Since \ac{FACSA} and \ac{FSCSA} share exactly the same stopping sets, these will be used in our evaluation. 

If we allow infinitely many decoding iterations (and let $\NRX \rightarrow \infty$ for \ac{FACSA}), all packet losses in CSA are caused by stopping sets. The PLR (see Definition~\ref{def:plr}) is then equivalent to the probability that an arbitrary VN is part of a stopping set.

We denote by $\A$ the set of all stopping sets and by $\Astar \subset \A$ a smaller set of minimal stopping sets that dominate the PLR in the EF region. Furthermore,  let $\user$ denote an arbitrary VN in a CSA system. The PLR can be approximated as follows, 
\begin{align}
	\PLR &= \Pr\left(\bigcup_{\Sset \in \A} \user \in \Sset \right)\stackrel{(a)}{\leq} \sum_{\Sset \in \A}\Pr\left(\user \in \Sset \right)  
	\stackrel{(b)}{\approx}\sum_{\Sset \in \Astar}\Pr\left(\user \in \Sset \right) \nonumber \\
	&\stackrel{(c)}{=}\sum_{\Sset \in \Astar}\sum_{m=0}^{\infty}\ps \Pr(\M=m).  \label{eq:PLR}
\end{align}
In $(a)$ the probability is upper bounded using the union bound. In $(b)$ we consider a summation over the subset $\Astar$, turning the upper bound into an approximation. Lastly, in $(c)$ we condition the probability of $u$ being part of a stopping set $\Sset$ on the RV $\M$, representing the number of VNs that can create the stopping set $\Sset$ with $\user$, and average over all possible values of $\M$.

Using \eqref{eq:PLR} as a starting point, we derive EF approximations for \ac{FACSAFNB} and \ac{FACSAUNB}. We do not consider boundary effects in order to simplify the analysis. Furthermore, we remark that a boundary has negligible impact on the  EF of a system that runs for a long time. 

We  express $\ps$ in \eqref{eq:PLR} in terms of factors that are simpler to derive,
\begin{equation}
	\ps =  \frac{\aS b(\Sset) \iso}{d(\Sset)} \cdot \frac{\numVNs}{m},
	\label{eq:PS}
\end{equation}
where $\aS$ is the expected number of ways to select $\numVNs$ VNs with the degree profile $\profileS$ from a set of $m$ VNs with degree distribution $\VNdd(x)$, $b(\Sset)$ is the number of ways to select the CNs of $\Sset$ such that $u \in \Sset$, 
$\iso$ is the number of graph-isomorphisms of $\Sset$,
and $d(\Sset)$ is the total number of ways in which $\numVNs$ VNs (including $u$) with degree profile $\profileS$ can connect edges to  CNs in their local frames. The fraction $\frac{\numVNs}{m}$ represents the probability that VN $u$ is one of the $\numVNs $  VNs in $\Sset$.

We give first the factor $a(\Sset)$, because it is the same for \ac{FACSAFNB} and \ac{FACSAUNB},
\begin{equation}
\aS = \binom{m}{\numVNs}  \numVNs! \prod_{l}^{}\frac{\VNdd_l^{\numVNsL}}{\numVNsL!},
\label{eq:aSFS}
\end{equation}
which stems from the multinomial distribution and was derived in \cite{ivanov2015broadcast}. In the following, we derive expressions for the factors $b(\Sset)$ and $d(\Sset)$.
\subsection{Frame Asynchronous CSA with First Slot Fixed}
Let $\user$ represent a VN in the range $[i,i+n-1]$. Furthermore, to simplify the derivation we make the assumption that $\Sset$ spans at most $n$ slots. Without loss of generality, we consider the range $[i,i+n-1]$. 

Since we are considering stopping sets constrained to the slots in the range $[i,i+n-1]$ that contain $u$, the first slot of the stopping set must be $i$. According to our assumption, the remaining $\numCNs-1$ slots of $\Sset$ are chosen with equal probability from the subsequent $\N-1$  slots. This gives
\begin{equation}
\bS \approx \binom{\N-1}{\numCNs-1},
\label{eq:bS-1st}
\end{equation}
where the subindex $\text{FA-F}$ is a short-hand notation indicating that the parameter, in this case $b$, is for \ac{FACSAFNB}. Similarly, we will use subindex $\text{FA-U}$ for \ac{FACSAUNB}.

We now consider $\dS$.  An arbitrary active user in slot $i+n-1$ has $\N$ equiprobable slots for its first replica, \IE the slots in $[i,i+n-1]$. However, the first replica of user $u$ is fixed to slot $i$. For each placement of a degree-$l$ user's first replica, there are $\binom{\N-1}{l-1}$ possible placements of its remaining replicas. Furthermore, each user places its replicas independently of other users. Thus,
\begin{equation}
\dS = n^{-1}\prod_{l}^{}\left(\N\binom{\N-1}{l-1}\right)^{\numVNsL}.
\label{eq:dS-1st}
\end{equation}

An approximation of the EF for \ac{FACSAF} is now given by evaluating \eqref{eq:PLR}, using \eqref{eq:PS}-\eqref{eq:dS-1st} and $\Pr(M=m)=e^{-ng}(ng)^m /m!$,
\begin{IEEEeqnarray*}{rCl}
& &\PLR_{\mathrm{FA-F}}\approx  \\
& & \sum_{\Sset \in \Astar}  \sum_{m=0}^{\infty}\frac{\aS\bS \iso}{\dS}\frac{\numVNs}{m} \frac{e^{-\N\G}(\N\G)^m}{m!} \IEEEyesnumber.
\label{eq:PLR_FA_F}
\end{IEEEeqnarray*}

\subsection{Frame Asynchronous CSA with Uniform Slot Selection}
We denote by $\user$ an arbitrary VN in an \ac{FACSAU} system. Without loss of generality, we assume that if a VN $\user \in \Sset$, then $\user$ is the highest degree VN of $\Sset$. We make a simplifying assumption that all VNs in $\Sset$ must be active in the entire range $[k_\mathrm{f},k_\mathrm{l}]$, where $k_\mathrm{f}$ and $k_\mathrm{l}$ are the positions of the first and last CNs that $\user$ is connected to, respectively, and we let  $\lmax$ denote the degree of $\user$.

If we denote by $D$ the RV representing the distance $k_\mathrm{l}-k_\mathrm{f}$, then its \ac{pmf} is given by
\begin{equation}
	\Pr(D=d) =  (n-d)\frac{\binom{d-1}{\lmax-2}}{\binom{n}{\lmax}},
	\label{eq:pmf_D}
\end{equation}
for $d \in [\lmax-1, n-1]$. According to our assumption, the number of VNs from which the VNs of $\Sset$ can be selected  is Poisson-distributed with mean $\G(\N-D)$. We let $M \sim \Po{\G(\N-D)}$ be the RV representing this number. Then, $m$ in \eqref{eq:PLR} is a realization of  $M$ such that
\begin{align}
&\Pr(M=m) = \nonumber \\ 
&\sum_{d=\lmax-1}^{n-1} \frac{e^{-\G(\N-d)}(\G(\N-d))^m}{m!}(n-d)\frac{\binom{d-1}{\lmax-2}}{\binom{n}{\lmax}} ,
\label{eq:PrMuni} 
\end{align}
obtained by averaging over $D$.

The CNs for $\Sset$ are selected randomly  from a set of $n$ CNs corresponding to the local frame of $u$. Therefore, 
\begin{equation}
\bSFAuni \approx \binom{n}{\numCNs}	.
\label{eq:bSFAuni}
\end{equation}
A degree-$l$ VN can connect its edges in $\binom{n}{l}$ ways to its local frame, hence,
\begin{equation}
\dSFAuni =  \prod_{l}^{} \binom{n}{l}^{\numVNsL}.
\label{eq:dSFAuni}
\end{equation}

Now, evaluating  \eqref{eq:PLR}, using \eqref{eq:aSFS}, \eqref{eq:bSFAuni}--\eqref{eq:dSFAuni}, and \eqref{eq:PrMuni} gives
\begin{align}
	\PLR_{\mathrm{FA-U}} \approx & \sum_{\Sset \in \Astar} \frac{\bSFAuni \iso}{\dSFAuni} \sum_{d= \lmax-1}^{n-1}\Pr(D=d) \nonumber  \\ 
  	                             & \sum_{m=0}^{\infty}\aSFS \frac{\numVNs}{m} \frac{e^{-\G(\N-d)}(\G(\N-d))^m}{m!}.	
	\label{eq:plrFAuni}
\end{align}

\subsection{Frame Synchronous CSA}
The probability $\ps$ in \eqref{eq:PS} for an FS system with constant number of users per frame $m$ has been previously derived in \cite{ivanov2015error} and \cite{ivanov2015broadcast}. For completeness, we give here the corresponding expressions of the factors in \eqref{eq:PS}, because the formulation that we use is slightly different and also includes the Poisson user model.

\begin{figure*}[!t]
	\vspace{-1ex}
	\begin{equation}\tag{51}
		\phi (\Sset)  = \begin{cases}
			\numCNs \prod_d d^{-\nu_d(\Sset)}\sum_{k=0}^{\numVNs-1} (-1)^{\numVNs-1+k} \frac{(\numVNs-1)!}{k!} (\N\G)^k & \text{for \ac{FACSAF}} \\
				\sum_{k=0}^{\numVNs-1} \sum_{d=\lmax-1}^{n-1} (-1)^{\numVNs-1+k} \frac{(\numVNs-1)!}{k!} ((\N-d)\G)^k  (n-d)\frac{\binom{d-1}{\lmax-2}}{\binom{n}{\lmax}} & \text{for \ac{FACSAU}}\\
			\sum_{k=0}^{\numVNs-1} (-1)^{\numVNs-1+k} \frac{(\numVNs-1)!}{k!} (\N\G)^k 								& \text{for \ac{FSCSA}} 
		\end{cases}
		\label{eq:phis}	
	\end{equation}
	\vspace{-1ex}
	\hrulefill
\end{figure*}

The CNs for $\Sset$ are selected randomly and uniformly from a set of $\N$ CNs (corresponding to the $n$ slots of the frame) and thus,
\begin{equation}
\bSFS = \binom{n}{\numCNs}	.
\label{eq:bSFS}
\end{equation}
A degree-$l$ VN can connect its edges in $\binom{n}{l}$ ways to the frame, hence,
\begin{equation}
\dSFS =  \prod_{l}^{} \binom{n}{l}^{\numVNsL}.
\label{eq:dSFS}
\end{equation}
Now, evaluating  \eqref{eq:PLR}, using \eqref{eq:aSFS}, \eqref{eq:bSFS}--\eqref{eq:dSFS}, and $\Pr(M=m)=e^{-ng}(ng)^m /m!$ gives
\begin{equation}
\PLR_{\mathrm{FS}} \approx  \sum_{\Sset \in \Astar}\sum_{m=0}^{\infty}\frac{\aSFS\bSFS \iso}{\dSFS}\frac{\numVNs}{m} \frac{e^{-\N\G}(\N\G)^m}{m!}.
\label{eq:PLR_FS}
\end{equation}

\subsection{Numerical Evaluation of the Error Floor Approximations}
We give an easy-to-use formula to evaluate \eqref{eq:PLR_FA_F}, \eqref{eq:plrFAuni}, and \eqref{eq:PLR_FS},
\begin{equation}
	\PLR \approx \sum_{\Sset \in \Astar}^{} 	\phi (\Sset)  \numVNs \iso \binom{n}{\numCNs} \prod_l \frac{\Lambda_l^{\nu_l(\Sset)}}{\nu_l(\Sset)!}\binom{n}{l}^{-\nu_l(\Sset)},
\end{equation}
where $\phi(\Sset)$ is given by \eqref{eq:phis} at the top of the next page.  In \eqref{eq:phis}, we used that
\setcounter{equation}{51}
\begin{align}
	&\sum_{m=0}^{\infty}\frac{e^{-ng}(ng)^m }{m(m-\numVNs)!}  = \nonumber \\
	&\sum_{k=0}^{\numVNs-1} (-1)^{\numVNs-1+k} \frac{(\numVNs-1)!}{k!} (\N\G)^k .
\end{align}
Thus, the infinite sum over $m$ in \eqref{eq:PLR_FA_F}, \eqref{eq:plrFAuni}, and \eqref{eq:PLR_FS} can be replaced by a finite sum. 

\section{Numerical Results}
\label{sec:results}
In this section, we give numerical results on the performance of \ac{FACSA} in the asymptotic and finite frame length regime and give comparisons with  \ac{FSCSA} and \ac{SCCSA} in terms of decoding thresholds, EF, and delay.

\subsection{Asymptotic Iterative Decoding Thresholds} 

\begin{table}[t]
	\addtolength{\tabcolsep}{-0.7mm}
	\scriptsize
	\caption{DE thresholds for FA-CSA and FS-CSA}
	\vspace{-3ex}
	\begin{center}\begin{tabular}{lccccccc}
			\hline
			\toprule
			$\VNdd(x)$     & $x^3$    & $x^4$   & $x^5$   & $x^6$ & $x^7$ & $x^8$ & $\VNddOpt(x)$\\
			\otoprule
	     	$\gbound$      & $0.940$  & $0.980$  & $0.993$ & $0.997$  & $0.999$  & $1.000$  & $0.973$\\[0.5mm]
			$\gstarfaBE$   & $0.917$  & $0.976$  & $0.992$ & $0.997$  & $0.998$  & $0.999$  & $0.963$\\[0.5mm]
			$\gstarfaUB$   & $0.917$  & $0.976$  & $0.992$ & $0.997$  & $0.998$  & $0.999$  & $0.963$\\[0.5mm]
			$\gstarfanoBE$ & $0.818$  & $0.772$  & $0.701$ & $0.637$  & $0.581$  & $0.534$  & $0.851$\\[0.5mm]
			$\gstarfanoUNB$& $0.818$  & $0.772$  & $0.701$ & $0.637$  & $0.581$  & $0.534$  & $0.851$\\[0.5mm]
			$\gstarfs$     & $0.818$  & $0.772$  & $0.701$ & $0.637$  & $0.581$  & $0.534$  & $0.851$\\[0.5mm]
			\bottomrule
		\end{tabular} \end{center}
		\label{tab:DEThreshold} 
		\vspace{-2ex}
	\end{table}
In Table~\ref{tab:DEThreshold}, we give iterative decoding thresholds for  \ac{FACSA}, for $\VNdd(x)=x^l$ with $l=$ 3, 4, 5, 6, 7, and 8, and $\VNdd(x)=\VNddOpt(x)=0.86x^3+0.14x^8$.  $\VNddOpt(x)$  was obtained in \cite{ivanov2015broadcast} for \ac{FSCSA} by a joint optimization of the EF and the threshold. {The upper bound on the achievable threshold according to \cite[(23)]{paolini2015csa}  for FS-CSA, denoted here by $\gbound$, is also given in Table~\ref{tab:DEThreshold}.}
\begin{figure}[]
	\centering
	\vspace{-3ex}
	\includegraphics[width=\linewidth]{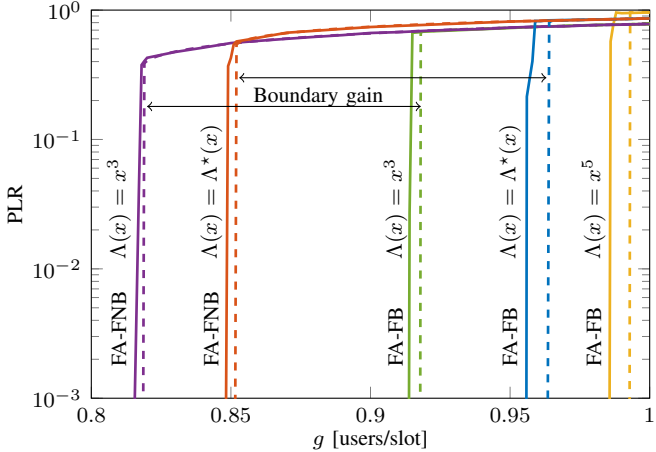}
	\vspace{-4ex}
	\caption{DE (dashed lines) and simulation results for $\N=10^5$ (solid lines) of the PLR for \ac{FACSAFB} and \ac{FACSAFNB}.}
	\label{fig:thresholds}
	\vspace{-3ex}
\end{figure} 

We observe that for \ac{FACSA} with boundary effect, the decoding threshold improves significantly with respect to the case where there are already active users at time $i=0$. This is due to a boundary effect (thus its name) caused by the lower degree of the CNs for $i \in [1,n-1]$, which results in a wave-like decoding effect similar to that of spatially coupled LDPC (SC-LDPC) codes \cite{Jim99,Kud11}. Furthermore, for \ac{FACSA} with boundary effect and regular VN degree distribution $\VNdd(x)=x^l$, the decoding threshold improves with increasing VN degree, whereas the opposite occurs for the systems without boundary effect. This behavior is similar to that of regular LDPC codes, where a larger VN degree improves the threshold for SC-LDPC codes but has the opposite effect for uncoupled LDPC codes. {Remarkably, increasing the VN degree the decoding threshold of FA-CSA approaches channel load $g=1$, i.e., that of perfect coordination. It is also interesting to observe that increasing the repetition factor allows to approach the bound in \cite{paolini2015csa} more tightly}. We also give in Table~\ref{tab:DEThreshold} the corresponding decoding  thresholds for FS-CSA, denoted by $\gstarfs$. \ac{FACSA} with boundary effect yields significantly better thresholds than \ac{FSCSA}. Interestingly, the thresholds for \ac{FACSAFNB}, \ac{FACSAUNB} and \ac{FSCSA} are identical. Indeed, the systems are very similar in that \ac{FSCSA} and \ac{FACSAUNB} have the same CN degree distribution and CNs of \ac{FACSAFNB} have the same average degree, but a slightly different node connectivity.

In \figref{thresholds}, we plot the PLR of FA-CSA with boundary effect obtained from DE (dashed lines) together with simulation results for $\N=10^5$ (solid lines), for  $\VNdd(x)=x^l$ with $l=3$ and $5$, and $\VNdd(x)=\VNddOpt(x)$. The figure shows that the DE equations are in good agreement with the simulations and make apparent the boundary gain for $\VNdd(x)=x^3$ and $\VNddOpt(x)$.

\begin{figure}[t!]
		\includegraphics[width=\linewidth]{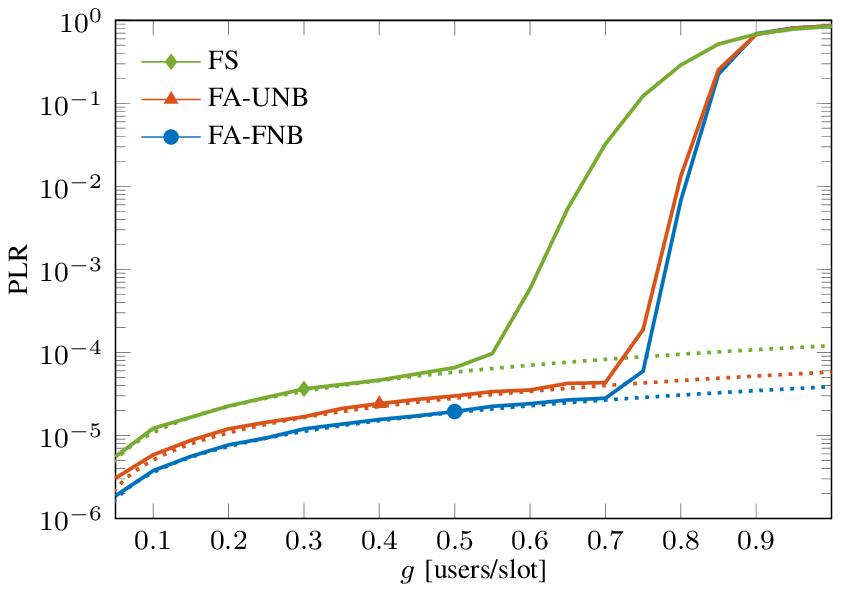}
		\vspace{-4.5ex}
		\caption{Simulated PLR (solid) and EF approximations (dotted) for \ac{FACSAFNB}, \ac{FACSAUNB} and \ac{FSCSA} with $n=200$ and $\VNdd(x) =\VNddOpt(x)$.}
		\label{fig:plr_all_200}
		\vspace{-3ex}
\end{figure}

\subsection{Finite Frame Length Packet Loss Rate and Error Floors}

In \figref{plr_all_200}, we plot the simulated PLR as a function of the system load, $g$, for \ac{FACSAFNB}, \ac{FACSAUNB}, and \ac{FSCSA}, with $\VNdd(x)=\VNddOpt(x)$ and  $\N=200$. The EF predictions, as derived in Section~\ref{sec:EF}, are also shown with dotted lines. We observe that both instances of  FA-CSA outperform FS-CSA in both the EF and the WF region. Furthermore, \ac{FACSAFNB} has a lower EF than \ac{FACSAUNB}, as predicted by the EF approximations. This hierarchy of the EF performance holds in general, \IE for any $\N$ and $\VNdd(x)$. We remark that FA-CSA with boundary effect (not included in the figure) exhibits the same performance in the EF as FA-CSA without boundary effect if the system runs for a long time.
Interestingly, despite the fact that \ac{FACSA} with no boundary effect has the same asymptotic decoding threshold as that of FS-CSA (see Table~\ref{tab:DEThreshold}), it is apparent from the figure that it exhibits much superior performance in the WF region as compared to \ac{FSCSA}. This seems to indicate that \ac{FACSA} without boundary effect has a better scaling of the PLR than \ac{FSCSA} in the finite frame length regime. 

\begin{figure}[t!]
		\vspace{-1.4ex}
		\includegraphics[width=\linewidth]{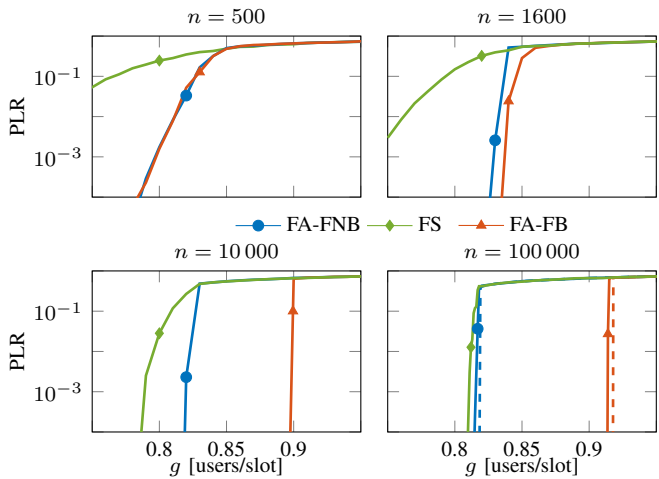}
		\vspace{-4ex}
		\caption{Simulated PLR performance in the WF region for \ac{FACSAFNB}, \ac{FACSAFB}, and \ac{FSCSA} with  $\VNdd(x)=x^3$  for increasing frame lengths $\N$.}
		\label{fig:finity_to_infinity}
		\vspace{-0ex}
\end{figure}

In \figref{finity_to_infinity}, we compare the PLR performance in the WF region of \ac{FACSAFB}, \ac{FACSAFNB}, and \ac{FSCSA} for  $\VNdd(x)=x^3$ and  frame lengths $n=500$, $1600$, $10\,000$, and $100\,000$.  For short frame lengths ($n=500$), \ac{FACSAFB} and \ac{FACSAFNB} have similar PLR performance, whereas \ac{FSCSA} performs worse. When the frame length is increased, however, the FA-CSA system with boundary effect outperforms the system without boundary effect in the WF region, as predicted by the DE thresholds. This is seen already for $n=1600$, for which the performance of \ac{FACSAFB} is slightly better as compared to that of \ac{FACSAFNB}. This \textit{boundary gain} increases with the frame length, as observed for $n=10\,000$ and $n=100\,000$. The asymptotic performance for \ac{FACSAFB} and \ac{FACSAFNB} given by DE is also plotted for $n=100\,000$ (dashed lines). As the frame length is increased, we also notice how the performance of \ac{FSCSA} approaches that of \ac{FACSAFNB}, as predicted by DE and the results in Table~\ref{tab:DEThreshold}.

We argued previously that the reason that systems with boundary effect yield better decoding thresholds is due to the fact that the lower degree CNs at the boundary induce a wave-like decoding effect. This improvement occurs only if the wave can propagate through the entirety of the system. Due to the randomness of the Poisson user model, for finite frame length the experienced load in a window of $n$ slots will sometimes be above and sometimes below the expected load $g$ users/slot. Such variations are more distinct for short frame lengths. Therefore, the reason that for short frame lengths the performance of FA-CSA with and without boundary effect are similar may be explained by the fact that the induced wave may be broken by events where the experienced load is large, causing the wave not to propagate further. Once the wave has been broken the FA-CSA system with boundary effect is equivalent to a system without boundary effect. For large frame lengths, instead, the variation of the experienced load is lower and the wave can propagate, improving performance in the WF region.

By careful inspection of \figref{finity_to_infinity}, we notice that the performance at system load $g \approx 0.83$ of \ac{FACSAFNB} is better for $n=1600$ compared to $n=100\,000$. This is counterintuitive at first, but might be explained by the same reasoning as above.  For particular frame lengths $n$, the variations of the experienced load could be large enough such that the low peaks with some chance will induce a decoding wave,  similar to the  wave of a system with boundary effect. To benefit from such events it would be necessary that the frame length $n$ is not too low, which would cause a decoding wave to break soon after its occurrence. If instead the frame length is large, the probability of a sporadic wave's creation would be extremely low due to the low variations of the experienced load. This conjecture is supported by simulation results.  For \ac{FACSAFNB} with $n=1600$ at a nominal system load $g \approx 0.83$,  we have observed a sudden drop of the simulated PLR from a high level down to the level of the EF after a number of slots. After the drop, the PLR remains low for the duration of the simulation, suggesting that a wave is propagating.  We remark  that this is indeed a sporadic behavior which occurs at different times (with respect to the start of the system) in each simulation round.

 \begin{figure}[t]
		\includegraphics[width=\linewidth]{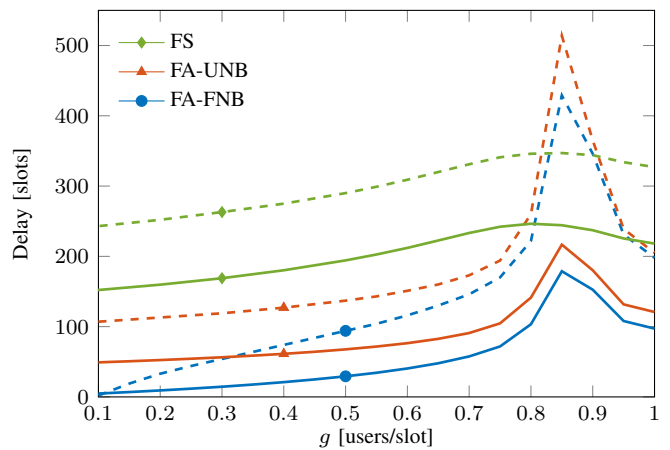} 		
 		\vspace{-3.5ex}
 		\caption{Delay performance of \ac{FACSAFNB}, \ac{FACSAUNB} and \ac{FSCSA}, with $\N=200$ and $\VNdd(x)=\VNddOpt(x)$. Solid lines show the average delay and dashed lines show the 90th percentile of the delay. }
 		\vspace{-2ex}
 		\label{fig:average_delay}
\end{figure}

\subsection{Finite Frame Length Delay Performance} 

We compare the delay (see Definition~\ref{def:delay}), of \ac{FACSAFNB}, \ac{FACSAUNB}, and \ac{FSCSA} for $\N=200$ and $\VNddOpt(x)$. For all schemes, we have considered a slot-by-slot decoding. We remark that some of the results presented here are not brand new, but confirm and further elaborate on the findings in \cite{meloni2012sliding}.

In \figref{average_delay}, the average and 90th percentile of the delay is depicted for $g \in [0.1,1]$. \ac{FACSAFNB} performs best in terms of average delay and \ac{FSCSA} worst. This is  expected  because a user in \ac{FACSAFNB} sends its replicas sooner after joining the system than in \ac{FACSAUNB} and \ac{FSCSA}. However, in terms of the 90th percentile, the two FA-CSA systems perform worse than \ac{FSCSA} for $\G \in [0.8,0.9]$. We remark that the delay is only defined for successfully received packets, and for $\G \in [0.8,0.9]$ the PLR is high for all three systems, as seen in \figref{plr_all_200}. Therefore, in practice, the system would not be operated at these loads.

 \begin{figure}[t]
		\includegraphics[width=\linewidth]{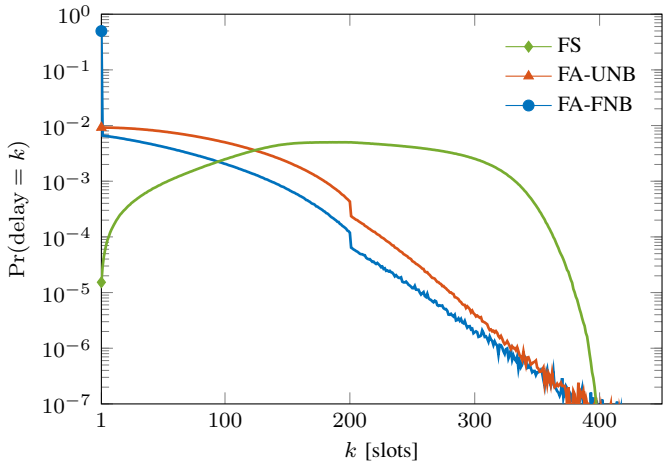}
 		\vspace{-4ex}
 		\caption{The \ac{pmf} of the delay for \ac{FACSAFNB}, \ac{FACSAUNB} and \ac{FSCSA}, with $\N=200$ and $\VNdd(x)=\VNddOpt(x)$ at a system load $g=0.5$. Dots mark the probability of a 1 slot delay.}
 		\vspace{-2ex}
 		\label{fig:pdf_delay}
 \end{figure}
 
\begin{figure}[t]
		\vspace{-1.7ex}
		\includegraphics[width=\linewidth]{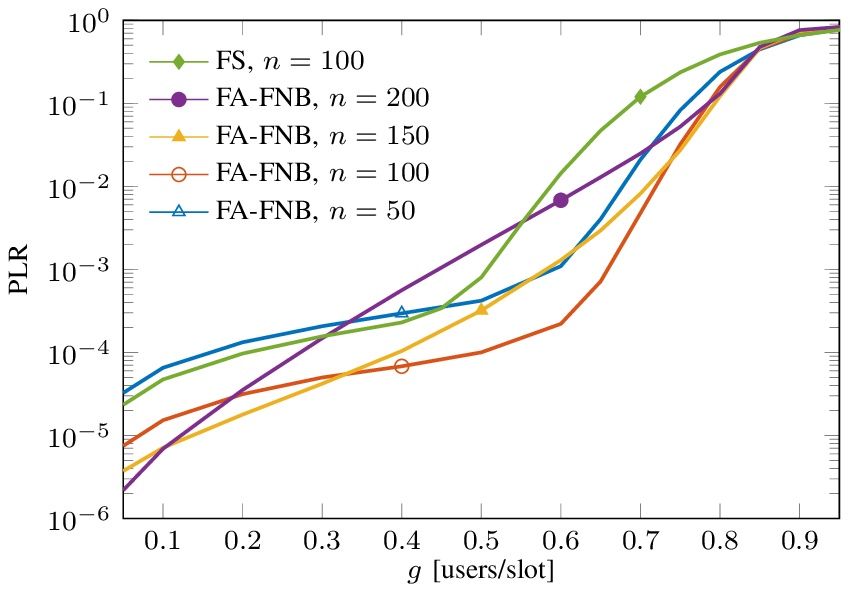}
		\vspace{-4ex}
		\caption{PLR performance with a maximum delay constraint $\delaymax=200$ for \ac{FACSAFNB} with varying local frame length $n$ and for \ac{FSCSA} with $n=100$, using $\VNdd(x)=\VNddOpt(x)$. }
		\vspace{-3ex}
		\label{fig:max_delay_200}
\end{figure}		
	
\begin{figure}[t]	
		\includegraphics[width=\linewidth]{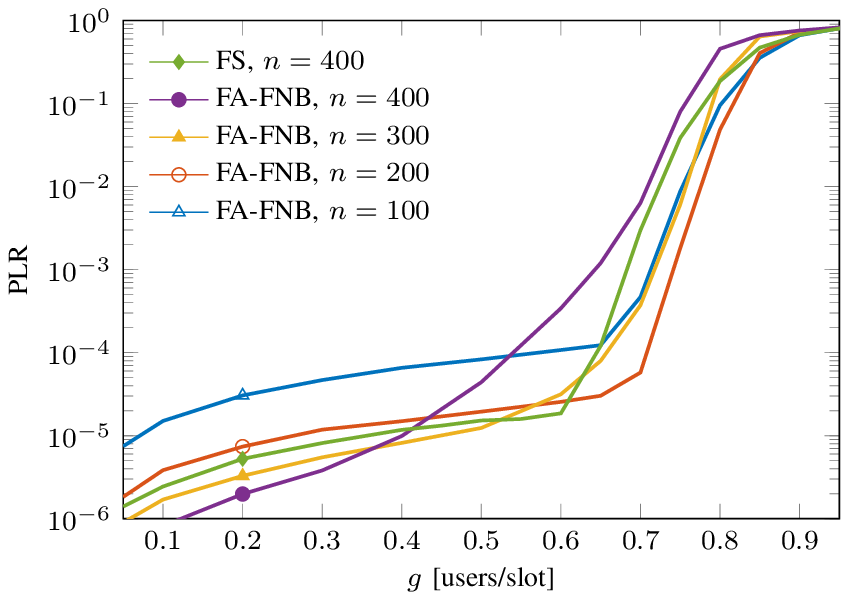}
		\vspace{-4.6ex}
		\caption{PLR performance with a maximum memory constraint $\NRX=400$ for \ac{FACSAFNB} with varying local frame length $n$ and for \ac{FSCSA} with $n=400$, using $\VNdd(x)=\VNddOpt(x)$.}
		\vspace{-2ex}
		\label{fig:max_complex_400}
\end{figure}

In \figref{pdf_delay}, we plot the \ac{pmf} of the delay, \IE the probability that a user has a certain delay $k$, for a system load $g=0.5$. The results show  again that, overall, \ac{FACSAFNB} provides the best delay performance. 
\figref{pdf_delay} also shows that the maximum delay of FA-CSA is larger than that of FS-CSA. However, the probability of such large delays is very low. In practice, the maximum delay of FA-CSA is limited by the frame length and the memory size of the receiver, whereas the maximum delay of FS-CSA is strictly limited by the frame length. Indeed, the maximum delay of \ac{FSCSA} is $2\N-1$, whereas the maximum delay of \ac{FACSA} is given by $\N+\NRX$. In \figreff{average_delay}{pdf_delay} we considered very large $\NRX$ in order  to not degrade the PLR performance. With a large $\NRX$, the maximum possible delay of \ac{FACSA} can be very large. For applications with  strict latency requirements this might be unacceptable. In \figreff{max_delay_200}{max_complex_400} we therefore present a comparison between \ac{FSCSA} and \ac{FACSAF} with a strict delay constraint of $\delaymax$ slots.   
\begin{definition}
	\label{def:plr_delay_constraint}
	The PLR of a CSA system with a delay constraint $\delaymax$, is the probability that the packet of an arbitrary user is not resolved within $\delaymax$ slots from the slot where the user joins the system.\footnote{Note that Definition~\ref{def:plr_delay_constraint} is not a constraint on the memory size $\NRX$.}
\end{definition}

In \figref{max_delay_200} we depict the delay-constrained PLR (according to Definition~\ref{def:plr_delay_constraint}) of \ac{FACSAFNB} with $\delaymax=200$ and compare it to that of \ac{FSCSA} with $\N= 100$, using $\VNdd(x) = \VNddOpt(x)$. We observe that for a given $\delaymax$ it is possible to find a local frame length for \ac{FACSAFNB} such that the PLR is strictly better than that of \ac{FSCSA} with the same delay constraint. A good choice of $\N$ for \ac{FACSAF} with the maximum delay constraint $\delaymax$ is, in general, half the length of the delay constraint, \IE $n=\delaymax/2$, as suggested by \figref{max_delay_200}. This choice provides relatively good performance and outperforms \ac{FSCSA} for all considered system loads.

A large memory can also be costly in practice. Therefore, in \figref{max_complex_400} we make a fair comparison of \ac{FACSAF} and \ac{FSCSA} in terms of memory size $\NRX$. In \ac{FSCSA} the only natural choice of memory size is the frame length $\N$, because a decoder gains nothing from capturing more than one frame simultaneously. For \ac{FACSA}, however, a fixed $\NRX$ leaves the choice of the local frame length $\N$ open. In the figure, we give PLR results for  \ac{FACSAFNB} with $\VNddOpt(x)$, $\NRX=400$, and different local frame lengths $\N$. We also plot the PLR of \ac{FSCSA} with $\N=400$. For almost all system loads, it is possible to find an appropriate $\N$ for \ac{FACSAF} so that it achieves better PLR compared to \ac{FSCSA} with the same memory constraint. Note that the advantage of \ac{FACSA} in terms of memory is that it is more flexible, \IE for a fixed memory length, the local frame length can be varied. If the memory size is not a constraint, the PLR performance of \ac{FACSA} is improved by increasing the memory size and adjusting the local frame length. 

\subsection{A Comparison with Spatially Coupled Coded Slotted ALOHA} 
\label{sec:SC}
\ac{SCCSA} is a frame-synchronous system where a degree-$l$ VN connects one edge to a randomly selected CN from each of $l$ consecutive frames \cite{liva2012spatially}. Furthermore, $w+l-1$  frames are grouped into a \textit{super-frame}.  The CNs of the $l-1$ first and last frames of the super-frame exhibit a lower average degree, creating a boundary effect in both ends of the super-frame.

\begin{figure}[t!]
		\includegraphics[width=\linewidth]{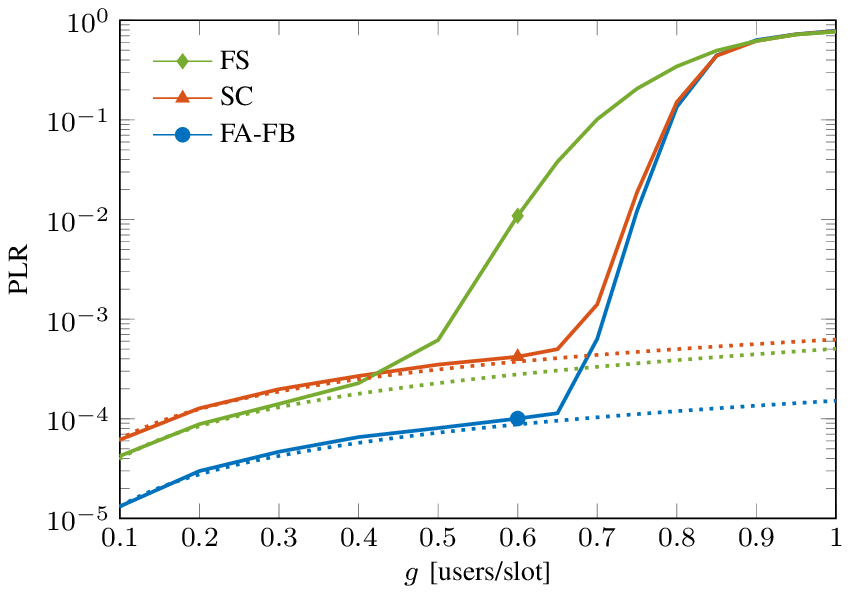}
		\vspace{-4ex}
		\caption{Simulated PLR (solid) and EF approximations (dotted) for \ac{FACSAFB}, \ac{FSCSA}, and \ac{SCCSA} with $n=120$ and $\VNdd(x)=x^3$.}
		\label{fig:sc_plr}
		\vspace{-3ex}
\end{figure}

In \cite[Table \RN{1}]{liva2012spatially} iterative decoding thresholds of \ac{SCCSA} were presented for $\VNdd(x)=x^l$ with $l=2,\,3,\,4,\,5$, and $6$. Surprisingly, the thresholds of \ac{SCCSA} are identical to the thresholds of \ac{FACSAFB} and \ac{FACSAUB} in Table~\ref{tab:DEThreshold}. This is remarkable because the systems are quite different. Indeed \ac{SCCSA} is more structured and enforces the spatially-coupled structure, whereas it is inherent to \ac{FACSA}. However, outside of the boundaries, the CN degree distributions of \ac{FACSA} and \ac{SCCSA} are identical (albeit a slightly different connectivity for \ac{FACSAFB}). Therefore, since both systems (with almost identical degree distributions) have a boundary that generates a wave-like decoding effect, similar decoding thresholds are expected.

Because of the similarities between \ac{SCCSA} and \ac{FACSA} with boundary effect, we present a comparison of \ac{FACSAFB}, \ac{SCCSA}, and \ac{FSCSA} in the finite frame length regime in terms of PLR and delay. In order to make a fair comparison of \ac{SCCSA}, \ac{FACSAFB}, and \ac{FSCSA}, we need to make some modifications to the system model for \ac{SCCSA} as it is described in \cite{liva2012spatially}. For the comparison, we consider that users join the system according to a slot-by-slot Poisson process, where $g$ is the expected number of users to join in a slot. Regular VN degree distributions are considered, \IE of the form $\VNdd(x)=x^l$. Furthermore, the frame length of the \ac{SCCSA} system is $n/l$. A user that joins an \ac{SCCSA} system will send one replica in each of the $l$ following frames, and we say that the user is active during these $l$ frames. This way, the largest span of the replicas of a user is equal for \ac{SCCSA}, \ac{FACSAFB}, and \ac{FSCSA}. Note that, by definition, the \ac{SCCSA} system assumes that there are no active users in the beginning, meaning that CNs of the first $l-1$ frames will exhibit lower expected degree than other CNs. We do not terminate the \ac{SCCSA} system which would cause the CNs of the last $l-1$ frames to have lower expected degree too, since this is not done for \ac{FACSAFB}. In practice, we can assume that the systems run indefinitely and a sliding-window decoder is used. Decoding of \ac{SCCSA} is performed in the same way as for \ac{FACSA}.

\begin{figure}[t!]
		\includegraphics[width=\linewidth]{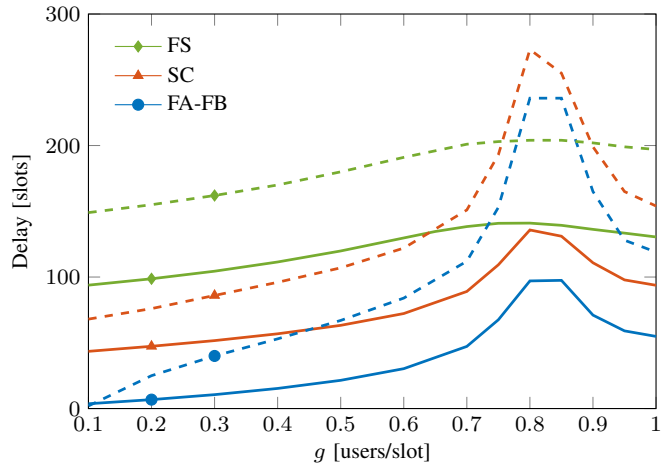}
		\vspace{-4ex}
		\caption{Simulated average delay (solid) and 90th percentile delay (dashed) for \ac{FACSAFB}, \ac{FSCSA}, and \ac{SCCSA} with $n=120$ and $\VNdd(x)=x^3$.}
		\label{fig:sc_delay}
		\vspace{-3ex}
\end{figure} 

\figref{sc_plr} gives simulation results on the PLR for \ac{FACSAFB}, \ac{SCCSA}, and \ac{FSCSA}, for $n=120$ and $l=3$. As expected, the WF performance of \ac{FACSAFB} and \ac{SCCSA} is similar. However, \ac{FACSAFB} performs remarkably better than \ac{SCCSA} in the EF. In fact, \ac{SCCSA} has even worse EF than \ac{FSCSA}.  The reason for this is that in \ac{SCCSA} each replica is forced into a smaller frame of size $n/l$. This makes the probability that two users select the same $l$ slots for transmission much larger.  In fact, this probability can be easily computed and used as an EF prediction for \ac{SCCSA} with a regular VN degree distribution $\VNdd(x)=x^l$,
\begin{equation}
	\PLR_{\mathrm{SC}} \approx  \sum_{m=0}^{\infty } m \left(\frac{l}{n}\right)^{l} \frac{e^{-g n/l}(g n /l)^m}{m!}=\left(\frac{l}{n}\right)^{l}\frac{gn}{l},
\end{equation}
which is plotted as the red dotted line in \figref{sc_plr}.

Additionally, the average and 90th percentile curves for the delay are given in \figref{sc_delay}.  For \ac{SCCSA} and \ac{FACSAFB}, the delay behavior is similar. Both systems allow a packet to be decoded after the reception of its last replica, which is why the 90th percentile delay is dramatically increased for loads corresponding to the WF-region of the PLR. However, this is not the region of interest, since it corresponds to a high PLR. Note that \ac{FACSAFB} yields better delay as compared to \ac{SCCSA}. This is due to the fact that for \ac{SCCSA} a user that joins the system needs to wait until the next frame before sending its first replica. Since the time a user in \ac{FSCSA} waits before its first replica is sent is even longer, the delay is even worse for \ac{FSCSA}. 

\section{Conclusions}
\label{sec:conclusion}
In this paper, we analyzed the asymptotic and finite frame length performance of frame asynchronous coded slotted ALOHA. We derived the DE that characterizes the asymptotic performance of FA-CSA and analytical approximations of its performance in the EF. If the receiver can monitor the system before users start transmitting or, equivalently, the system can be reinitialized, a boundary effect similar to that of spatial coupling appears, which greatly improves the decoding threshold as compared to that of standard FS-CSA. We showed that for finite frame length, \ac{FACSA} with and without boundary effect achieves superior performance than FS-CSA in terms of EF, WF, and delay. 
Furthermore, we compared the PLR of \ac{FACSA} and \ac{FSCSA} with constraints on both the maximum allowed delay and the memory size. \ac{FACSA} is more flexible in terms of frame length and memory size, and can typically be adjusted to outperform \ac{FSCSA}.
Additionally, we showed that \ac{FACSA} performs better than \ac{SCCSA} in the finite frame length regime, both in terms of EF and delay.


\ifCLASSOPTIONcaptionsoff
  \newpage
\fi
\bibliographystyle{IEEEtran}

%



\end{document}